\documentclass[prl,twocolumn,superscriptaddress,showpacs]{revtex4}

\usepackage{amsmath}
\usepackage{amssymb}
\usepackage{graphicx}
\usepackage{subfigure}

\newcommand{\be}[1]{\begin{equation}\label{#1}}
\newcommand{\ee}{\end{equation}}
\newcommand{\ba}[1]{\begin{eqnarray}\label{#1}}
\newcommand{\ea}{\end{eqnarray}}

\newcommand{\R}{ \mbox{\rm I$\!$R} }

\begin{document}

\title{Paradoxical transitions to instabilities in hydromagnetic Couette-Taylor flows}

\author{Oleg N. Kirillov}
%\email{o.kirillov@hzdr.de}
\affiliation{Helmholtz-Zentrum Dresden-Rossendorf,
P.O. Box 510119, D-01314 Dresden, Germany}

\author{Dmitry E. Pelinovsky}
%\email{dmpeli@math.mcmaster.ca}
\affiliation{Department of Mathematics, McMaster University,
Hamilton, Ontario, Canada, L8S 4K1}

\author{Guido Schneider}
%\email{guidos@mathematik.uni-stuttgart.de}
\affiliation{Institute of Analysis, Dynamics, and Modeling,
University of Stuttgart, Stuttgart, Germany}

\date{\today}

\begin{abstract}
By methods of modern spectral analysis, we rigorously find distributions of eigenvalues
of linearized operators associated with an ideal
hydromagnetic Couette-Taylor flow. Transition to instability
in the limit of vanishing magnetic field has a discontinuous change
compared to the Rayleigh stability criterion for hydrodynamical flows,
which is known as the Velikhov-Chandrasekhar paradox.
\end{abstract}

\pacs{02.30.Tb, 46.15.Ff, 47.32.-y, 47.85.L-, 47.35.Tv, 47.65.-d, 97.10.Gz, 95.30.Qd }

\maketitle

Instabilities of Couette-Taylor (CT) flow between two rotating cylinders
are at the cornerstone of the last-century hydrodynamics \cite{CI94}.
In 1917 Rayleigh found a necessary and sufficient condition for centrifugal instability of CT-flow of
an ideal fluid between cylinders of infinite length with respect to axisymmetric perturbations \cite{r17}.
Taylor extended Rayleigh's result to viscous CT-flow and computed seminal linear stability diagrams
that perfectly agreed with the experiment at moderate angular velocities \cite{t23}.

Despite the theoretical and experimental studies of the Couette-Taylor flow are more than a century long,
recent decade had seen a true renaissance of this classical subject caused by the increased demands of the
actively developing laboratory experiments with liquid metals that rotate in an external magnetic field \cite{lathrop04}.
The prevalence of resistive dissipation over viscous dissipation in liquid metals
dictates unprecedentedly high values of the Reynolds number (${\rm Re}\sim 10^6$) at the threshold
of the \textit{magnetorotational instability} (MRI) of hydrodynamically
stable quasi-Keplerian flows that currently is considered as the most probable trigger of
turbulence in astrophysical accretion discs  \cite{bh91}. Difficulties in keeping hydrodynamical
CT-flows laminar at such high speeds, put the laboratory detection of MRI at the edge of modern
technical capabilities.

Is the existing theory of MRI well-prepared to face these promising experimental opportunities?
No matter how paradoxical it may sound, the answer is: Not yet.

Indeed, already the discoverers of MRI, Velikhov \cite{v59} and Chandrasekhar \cite{c60},
pointed out a counter-intuitive phenomenon.
In case of an ideal non-resistive flow, which we
consider in this Letter, boundaries of the region of the magnetorotational instability
are misplaced compared to the Rayleigh boundaries of the region of the centrifugal
instability and do not converge
to those in the limit of negligibly small axial magnetic field.
In presence of dissipation the convergence is possible \cite{DiPrima84}.

The existing attempts of the physical explanation of the
\textit{Velikhov--Chandrasekhar paradox} \cite{ah73} involve
{ Alfv$\rm \acute e$n}'s theorem that `attaches' magnetic field lines
to the fluid of zero electrical resistivity, independent of the strength of
the magnetic field, which implies conservation of the angular velocity
(Velikhov-Chandrasekhar) rather than the angular momentum (Rayleigh).
However, the weak point of this argument is that the actual boundary of
MRI does depend on the magnetic field strength even in the case of ideal
MHD and tends to that of solid body rotation only when the field is
vanishing.  This indicates that the roots of the paradox are hidden deeper.

Recently, this intriguing effect was reconsidered in the full viscous and
resistive setting by a local WKB approximation \cite{ks11}.
It was found that the threshold surface of MRI in the space of resistive frequency,
{Alfv$\rm \acute e$n} frequency and Rossby number possesses a structurally stable
singularity known as the \textit{Pl\"ucker conoid} that persists at any level of viscous dissipation.
The singular surface connects the Rayleigh- and the Velikhov-Chandrasekhar thresholds
through the continuum of intermediate states parameterized by the Lundquist number \cite{ks11}.

Why does this singularity exist? Our Letter sheds light to this question
via rigorous inspection of the spectra of the boundary eigenvalue problems
associated with the ideal hydrodynamic and hydromagnetic CT-flows.
Rigorous spectral results are illustrated by MATLAB computations of
eigenvalues of the linearized operators.

If ${\bf u}$ is the velocity field, ${\bf b}$ is the magnetic field, and cylindrical coordinates
$(r,\theta,z)$ are used, the basic CT-flow
between cylinders of radii $R_1$ and $R_2$, $R_1 < R_2$, is
\begin{equation}
\label{basic-flow}
{\bf u}_0 = r \Omega(r) {\bf e}_{\theta}, \quad
{\bf b}_0 = b_0 {\bf e}_z, \quad \Omega(r) = a + {c}{r^{-2}},
\end{equation}
where $b_0$ is arbitrary and $(a,c)$ are related uniquely to
$\Omega_{1,2} = \Omega(R_{1,2})$ through the viscous limit,
\begin{equation}
\label{constants-a-b}
a = \frac{\Omega_2 R_2^2 - \Omega_1 R_1^2}{R_2^2 - R_1^2}, \quad
c = \frac{(\Omega_1 - \Omega_2) R_1^2 R_2^2}{R_2^2 - R_1^2}.
\end{equation}
In the case of co-rotating cylinders, $\Omega_{1,2} > 0$,
the Rayleigh boundary corresponds to $a = 0$, whereas
the Velikhov-Chandrasekhar boundary is $c = 0$.

The summary of our results is as follows.

(I) In the case of no magnetic field ($b_0 = 0$), co-rotating cylinders
($\Omega_{1,2} > 0$), and an ideal fluid, we prove that the linearized stability problem has
a countable set of neutrally stable pairs of (purely imaginary) eigenvalues
for $a > 0$ and a set of unstable pairs of (purely real) eigenvalues for $a < 0$,
all accumulating to zero. At $a = 0$, all pairs of eigenvalues merge together at zero.

(II) Under the same conditions but for counter-rotating cylinders with $\Omega_1 < 0$ and $\Omega_2 > 0$,
we show that there exist two sets of eigenvalue pairs: one set contains real eigenvalues and
the other set contains purely imaginary eigenvalues. The unstable real eigenvalues converge to
the zero accumulation point when $\Omega_1 \to 0$ for fixed $\Omega_2 > 0$ (where $a > 0$), whereas
the stable imaginary eigenvalues persist across $\Omega_1 =0$.

(III) For any magnetic field ($b_0 \neq 0$), co-rotating cylinders
($\Omega_{1,2} > 0$),  and an ideal non-resistive hydromagnetic flow, we prove that
there exist two sets of eigenvalue pairs and both sets contain
only purely imaginary eigenvalues for $0 < \Omega_1 < \Omega_2$. One set
remains purely imaginary for $\Omega_1 > \Omega_2$ but the other set
transforms to the set of real eigenvalues along a countable sequence of curves,
which are located for $\Omega_1 > \Omega_2$ and approach the diagonal line $\Omega_1 = \Omega_2$
($c = 0$) in the limit $b_0 \to 0$.
One pair of purely imaginary eigenvalues below the corresponding
curve transforms into a pair of unstable real eigenvalues above the curve. No eigenvalues
pass through the origin of the complex plane in the neighborhood of the line $a = 0$,
even if $b_0$ is close to zero.

(IV) Under the same conditions but for counter-rotating cylinders with $\Omega_1 < 0$ and $\Omega_2 > 0$,
we show the existence of four sets of eigenvalue pairs, which are either
purely imaginary or real. The unstable eigenvalues bifurcate again along
a countable sequence of curves, which are located for $\Omega_1 < 0$ and
approach $\Omega_1 = 0$ in the limit $b_0 \to 0$. The purely imaginary pair of eigenvalues
above the curve turns into a purely real pair of eigenvalues below the curve.

Although the results (I) and (II) partially reproduce the conclusions of Synge \cite{s33},
the existence of zero eigenvalues of infinite multiplicity at the Rayleigh threshold is emphasized here for the first time.
Similar coalescence of all eigenvalues at the zero value happens also in the Bose-Hubbard dimer %in the limit of vanishing interaction between the particles
\cite{eva08}.
Results (III) and (IV) are new to the best of our knowledge. Numerical evidences of
these results can be found in \cite{kcg02}.

The rest of our paper is devoted to the proofs of the above results and their numerical illustrations.
We take the equations for an ideal hydromagnetic fluid \cite{ah73},
\begin{equation}
\label{NS-equations}
\left. \begin{array}{l} {\bf u}_t + ({\bf u} \cdot \nabla) {\bf u} = - \nabla \left( p + \frac{1}{2} |{\bf b}|^2 \right)
+ ({\bf b} \cdot \nabla) {\bf b}, \\
{\bf b}_t = \nabla \times ( {\bf u} \times {\bf b}), \\
\nabla \cdot {\bf u} = 0, \quad
\nabla \cdot {\bf b} = 0, \end{array} \right\}
\end{equation}
where $p$ is the pressure term determined from the incompressibility condition $\nabla \cdot {\bf u} = 0$.
We linearize (\ref{NS-equations}) at the basic flow (\ref{basic-flow}) and use the standard separation of variables for
symmetric ($\theta$-independent) perturbations,
\begin{equation}
\label{separation}
{\bf u} = {\bf u}_0 + {\bf U}(r) e^{\gamma t + i k z}, \quad
{\bf b} = {\bf b}_0 + {\bf B}(r) e^{\gamma t + i k z},
\end{equation}
where $\gamma$ is the growth rate of perturbations in time and $k \in \mathbb{R}$ is the
Fourier wave number with respect to the cylindrical coordinate $z$. Performing routine
calculations \cite{DiPrima84}, we find the system of four coupled equations for components of ${\bf U}$ and ${\bf B}$ in the directions
of ${\bf e}_r$ and ${\bf e}_{\theta}$ (denoted by $U_r$, $U_{\theta}$, $B_r$, and $B_{\theta}$),
\begin{equation}
\label{linear-equations}
\left. \begin{array}{l} i k b_0 (k^2 + L) B_r + 2 k^2 \Omega(r) U_{\theta} = \gamma (k^2 + L) U_r, \\
i k b_0 B_{\theta} - 2 a U_r = \gamma U_{\theta}, \\
i k b_0 U_r = \gamma B_r, \\
i k b_0 U_{\theta} - \frac{2 c}{r^2} B_r = \gamma B_{\theta}, \end{array} \right\}
\end{equation}
where $L = -\partial_r^2 - \frac{1}{r} \partial_r + \frac{1}{r^2}$ is the Bessel operator,
which is strictly positive and self-adjoint with respect to the weighted inner product $\langle f, g \rangle =
\int_{R_1}^{R_2} r f(r) g(r) dr$. We note that $z$-components of ${\bf U}$ and ${\bf B}$, as well as the pressure term,
have been eliminated from the system of equations (\ref{linear-equations}) under the condition $k \neq 0$.

For hydrodynamic instabilities of the CT-flow, we set $b_0 = 0$, which yields
uniquely $B_r = B_{\theta} = 0$, $2 a U_r + \gamma U_{\theta} = 0$, and a closed
linear eigenvalue problem,
\begin{equation}
\label{eigenvalueRT}
\gamma^2 (k^2 + L) U_r = - 4 k^2 a \Omega(r) U_r, \quad R_1 < r < R_2,
\end{equation}
subject to the Dirichlet boundary conditions at the inner and outer cylinders $U_r(R_1) = U_r(R_2) =0$.

The operator $L$ is an unbounded strictly positive operator with a purely discrete spectrum of
positive eigenvalues $\{ \mu_n \}_{n \in \mathbb{N}}$ that diverge to infinity according
to the distribution $\mu_n \sim n^2$ as $n \to \infty$. Inverting this operator
for any $k \in \R$ and defining a new eigenfunction $\Psi$ by $U_r = (k^2 + L)^{-1/2} \Psi$,
we rewrite (\ref{eigenvalueRT}) in the form,
\begin{equation}
\label{T}
\gamma^2 \Psi = -a T \Psi, \quad T = 4 k^2 (k^2 + L)^{-1/2} \Omega (k^2 + L)^{-1/2},
\end{equation}
where the self-adjoint compact operator $T$ has eigenvalues $\{ -\gamma^2/a \}_{n \in \mathbb{N}}$
that accumulate to zero with $\gamma_n = {\cal O}(n^{-1})$ as $n \to \infty$.

If $\Omega_1, \Omega_2 > 0$, then $\Omega(r) > 0$ for all $r \in [R_1,R_2]$ and $T$ is a compact positive
operator. Hence, all $\gamma^2_n < 0$ if $a > 0$ and all $\gamma^2_n > 0$ if $a < 0$. The condition $a = 0$
($\Omega_2 R_2^2 = \Omega_1 R_1^2$) is the Rayleigh boundary, at which all
eigenvalues are at $\gamma = 0$. The proof of (I) is complete.

If $\Omega_1 < 0$ and $\Omega_2 > 0$, then $a > 0$ but $\Omega$ is sign-indefinite on $[R_1,R_2]$. Since $T$ is a compact
sign-indefinite operator, it has two sequences of eigenvalues accumulating to zero: one sequence has $\gamma^2_n < 0$
and the other one has $\gamma^2_n > 0$. This completes the
proof of (II).

\begin{figure}
\centering
\subfigure{(a)}{\includegraphics[width=0.5\textwidth]{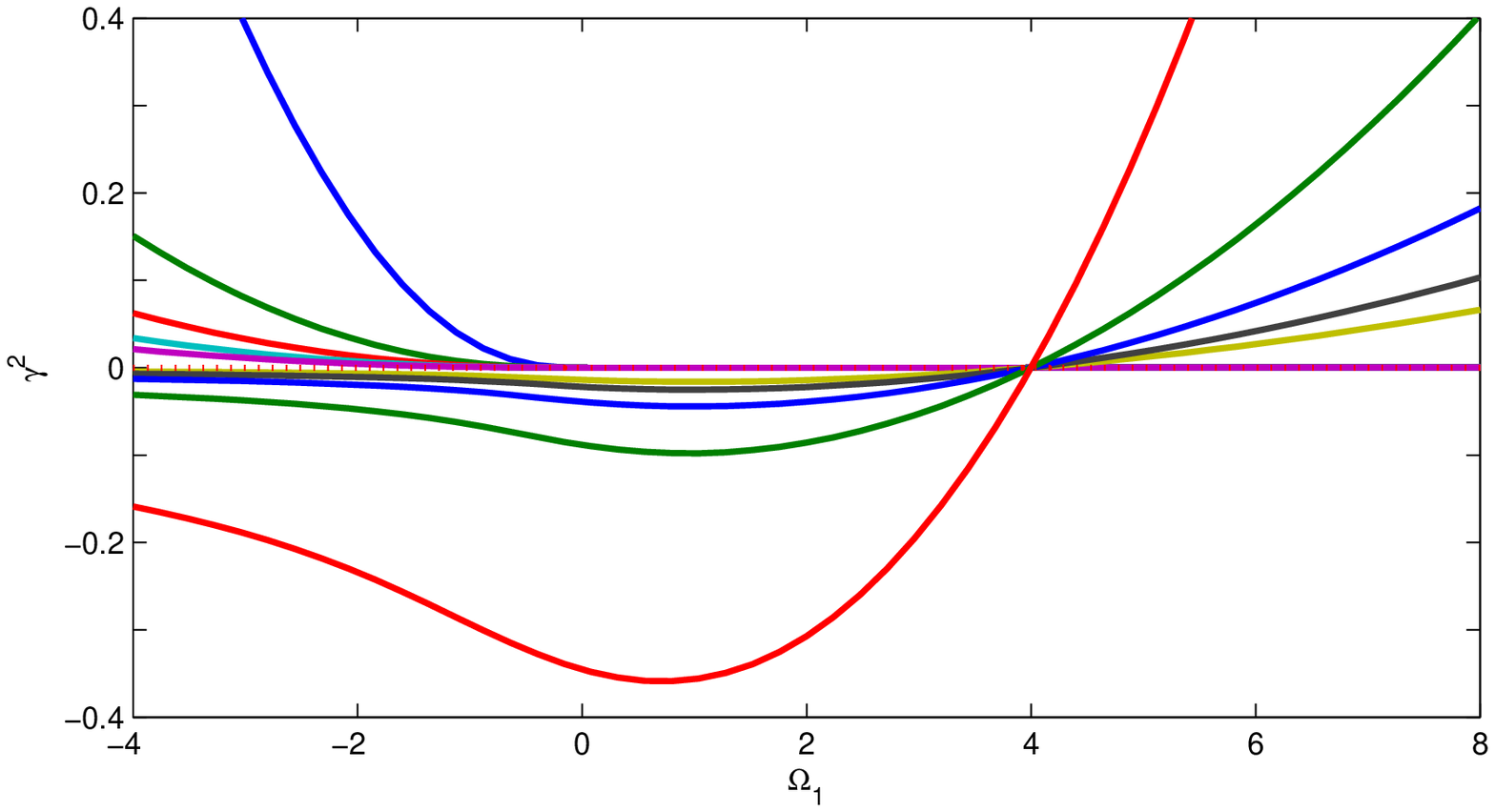}}
\subfigure{(b)}{\includegraphics[width=0.5\textwidth]{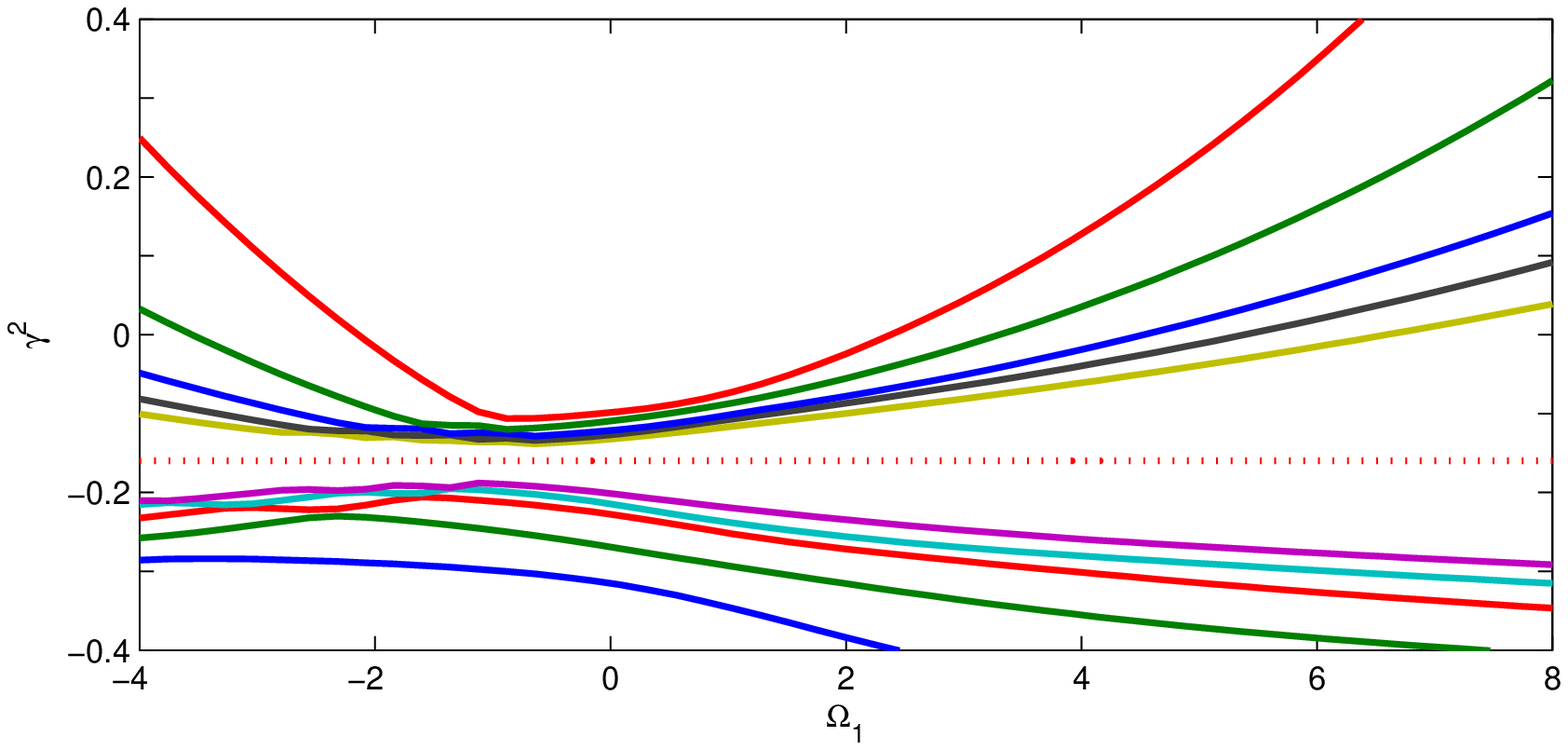}}
\caption{Squared eigenvalues $\gamma^2$ of the problem (\ref{eigenvalueRT}) versus $\Omega_1$ for $\Omega_2 = 1$, $R_1 = 1$, $R_2 = 2$, and $k = 1$.
(a) $b_0=0$: all squared eigenvalues $\gamma$ coalesce to zero at the Rayleigh line $\Omega_1 R_1^2 = \Omega_2 R_2^2$,
whereas positive squared eigenvalues for $\Omega_1 < 0$ merge to zero at $\Omega_1 = 0$. (b) $b_0=0.4$:
the squared eigenvalues of the problem (\ref{eigenvalueCV}) change stability above the Velikhov-Chandrasekhar line $\Omega_1 = \Omega_2$ and
below $\Omega_1 = 0$.}
\label{Figure1}
\end{figure}

Figure \ref{Figure1}(a) gives numerical approximations of the five largest and five smallest
squared eigenvalues $\gamma^2$ as functions of the parameter $\Omega_1$
for fixed values of $\Omega_2 = 1$, $R_1 = 1$, $R_2 = 2$, and $k = 1$. The
dotted line shows the accumulation point $\gamma = 0$ for the sequences of eigenvalues.
For $\Omega_1 > 0$, the five smallest eigenvalues are not distinguished from
the zero accumulation point.

For hydromagnetic instabilities, we express $B_r$, $B_{\theta}$, and $U_{\theta}$ from the
system of linearized equations (\ref{linear-equations}) and find a closed linear eigenvalue problem,
\begin{equation}
\label{eigenvalueCV}
(\gamma^2 + k^2 b_0^2)^2 (k^2 + L) U_r = 4 k^2 \Omega(r) \left( \frac{k^2 b_0^2 c}{r^2} - a \gamma^2 \right) U_r,
\end{equation}
subject to the same Dirichlet boundary conditions at $r = R_{1,2}$. If $b_0 = 0$ and
$\gamma \neq 0$, system (\ref{eigenvalueCV})
reduces to (\ref{eigenvalueRT}), however, it is a bi-quadratic eigenvalue problem
and hence has a double set of eigenvalues compared to (\ref{eigenvalueRT}).

Denoting $\lambda = \gamma^2 + k^2 b_0^2$, we rewrite (\ref{eigenvalueCV}) as
the quadratic eigenvalue problem,
\begin{equation}
\label{eigenvalueCVquad}
\lambda^2 (k^2 + L) U_r + 4 a k^2 \lambda \Omega(r) U_r = 4 k^4 b_0^2\Omega^2(r) U_r.
\end{equation}
It follows again from the compactness of the operators $(k^2 + L)^{-1} \Omega$ and
$(k^2 + L)^{-1} \Omega^2$ that the spectrum of the quadratic eigenvalue problem (\ref{eigenvalueCVquad})
is purely discrete. Chandrasekhar \cite{c60} showed that all eigenvalues $\lambda$ are real.
 We shall prove that these eigenvalues accumulate to zero as two countable
sets with $\lambda_n = \mathcal{O}(n^{-1})$ as $n \to \infty$, one set is
for positive $\lambda$ and the other set is for negative $\lambda$.
The result definitely holds for $a = 0$ because $\lambda^2$ becomes an eigenvalue
of the self-adjoint problem,
\begin{equation}
\label{S}
\lambda^2 \Psi = k^2 b_0^2 S \Psi, \quad
S = 4 k^2 (k^2 + L)^{-1/2} \Omega^2 (k^2 + L)^{-1/2},
\end{equation}
where $S$ is a compact positive operator.

To show the same conclusion for $a \neq 0$, we use a recently developed technique
from \cite{Kollar} and rewrite (\ref{eigenvalueCVquad}) as a parameter continuation problem
for $\nu = \lambda^{-1}$,
\begin{equation}
\label{eigenvalueCVpar}
a \nu \Omega(r) U_r = - \frac{1}{4 k^2} (k^2 + L) U_r + k^2 b_0^2 \epsilon^2 \Omega^2(r) U_r.
\end{equation}
Here eigenvalues $\nu$ of (\ref{eigenvalueCVpar}) for $a \neq 0$ are continued with respect
to the real values of $\epsilon$ to recover eigenvalues $\lambda = \nu^{-1}$
of (\ref{eigenvalueCVquad}) at the intersections with the diagonal $\nu = \epsilon$.

At $\epsilon = 0$, we recover back the hydrodynamical problem (\ref{eigenvalueRT}).
If $\Omega_1, \Omega_2 > 0$, then $\Omega(r) > 0$ for all $r \in [R_1,R_2]$ and
eigenvalues $\{ \nu_n(\epsilon) \}_{n \in\mathbb{N}}$ at $\epsilon = 0$ are strictly negative if $a > 0$ or
strictly positive if $a < 0$. Moreover, $\nu_n(0) \sim n^2$ as $n \to \infty$. Without loss of
generality, let us consider the case $a > 0$. Each negative eigenvalue $\nu_n(\epsilon)$
is strictly increasing for large values of $|\epsilon|$ at any point $\epsilon_0$, because
\begin{equation}
\label{derivative}
a \epsilon_0 \frac{d \nu_n}{d \epsilon} \biggr|_{\epsilon = \epsilon_0} = 2 \epsilon_0^2 k^2 b_0^2
\frac{\langle \Omega^2 \varphi_n, \varphi_n \rangle}{\langle \Omega \varphi_n, \varphi_n \rangle} > 0,
\end{equation}
where $\varphi_n$ is the eigenfunction for the eigenvalue $\nu_n(\epsilon)$
in (\ref{eigenvalueCVpar}) at $\epsilon = \epsilon_0$.
The right-hand-side of (\ref{derivative}) is always bounded, hence the eigenvalues $\{\nu_n(\epsilon) \}_{n \in \mathbb{N}}$
are continued to positive infinity as $|\epsilon| \to \infty$. As a result, there exist two countable sets of
intersections of eigenvalues $\{\nu_n(\epsilon) \}_{n \in \mathbb{N}}$  with $\nu = \epsilon$,
one set is for positive $\lambda = \nu^{-1}$ and the other set is for negative $\lambda$.
Both sets accumulate at zero as $n \to \infty$. This completes the
proof of (III).

If $\Omega_1 < 0$ and $\Omega_2 > 0$, then $a > 0$ but $\Omega$ is sign-indefinite on $[R_1,R_2]$.
In this case, again using the compact operator $T$ in (\ref{T}),
there exist two sets of eigenvalues $\{ \nu_n^{\pm}(\epsilon) \}_{n \in\mathbb{N}}$ of (\ref{eigenvalueCVpar}) at $\epsilon = 0$:
one set $\{ \nu_n^-(0) \}_{n \in\mathbb{N}}$ is strictly negative with $\langle \Omega \varphi_n^-, \varphi_n^- \rangle < 0$ and
the other set $\{ \nu_n^+(0) \}_{n \in\mathbb{N}}$ is strictly positive with $\langle \Omega \varphi_n^+, \varphi_n^+ \rangle > 0$.
Because the signs of $\langle \Omega \varphi_n^{\pm}, \varphi_n^{\pm} \rangle$ are preserved for small $\epsilon \neq 0$,
it follows from the derivative (\ref{derivative}) that the eigenvalues $\{ \nu_n^-(\epsilon) \}_{n \in\mathbb{N}}$ are convex
upward for larger values of $|\epsilon|$ and the eigenvalues $\{ \nu_n^+(\epsilon) \}_{n \in\mathbb{N}}$ are concave downward for larger values
of $\epsilon$. The curves of $\{ \nu_n^{\pm}(\epsilon) \}_{n \in\mathbb{N}}$ may intersect but the intersection is
safe (i.e., eigenvalues split without onset of complex eigenvalues) because the eigenvalue problem (\ref{eigenvalueCVpar})
is self-adjoint for any real $\epsilon$ and hence multiple eigenvalues are always semi-simple.
If the signs of $\langle \Omega \varphi_n^{\pm}, \varphi_n^{\pm} \rangle$ are preserved along the entire
curves, then we conclude on the existence of four sets of intersections of these eigenvalues with the main
diagonal $\nu = \epsilon$: two sets give positive eigenvalues $\lambda$ and the two other sets give negative eigenvalues.
The conclusion is not affected by the fact that $\langle \Omega \varphi_n^{\pm}, \varphi_n^{\pm} \rangle$ may vanish
along the curve. If this is happened, then $\langle \Omega \varphi_n^{\pm}, \varphi_n^{\pm} \rangle$ has
at least a simple zero due to analyticity in $\epsilon$ and hence the derivative (\ref{derivative}) implies
that the corresponding curve $\nu_n^{\pm}(\epsilon)$ goes to plus or minus infinity for finite values of $\epsilon$.
This argument completes the proof of (IV).

Figure \ref{Figure1}(b) shows numerical approximations of the five smallest and five largest
squared eigenvalues $\gamma^2$ as functions of $\Omega_1$ for fixed values of $\Omega_2 = 1$,
$R_1 = 1$, $R_2 = 2$, $b_0 = 0.4$, and $k = 1$. Cascades of instabilities arise for $\Omega_1 > \Omega_2$
and $\Omega_1 < 0$ by subsequent merging of pairs of purely imaginary eigenvalues $\gamma$
at the origin and splitting into pairs of real (unstable) eigenvalues $\gamma$.
For $\Omega_1 > 0$, the two sets of squared eigenvalues
accumulate to the value $\gamma^2 = -k^2 b_0^2$ ($\lambda = 0$), which is shown by the dotted line.
For $\Omega_1 < 0$, a more complicated behavior is observed within each set: the squared eigenvalues coalesce
and split safely, indicating that each set is actually represented by two disjoint sets
of the squared eigenvalues.

To study the instability boundaries in (\ref{eigenvalueCV}), we substitute $\gamma = 0$
and regroup terms for $b_0 \neq 0$ to obtain
\begin{equation}
\label{eigenvalueCVzero}
b_0^2 (k^2 + L) U_r = 4 (\Omega_1 - \Omega_2) \frac{R_1^2 R_2^2 \Omega(r)}{(R_2^2 - R_1^2) r^2} U_r,
\end{equation}

If $\Omega_{1,2} > 0$, it follows from equation (\ref{eigenvalueCVzero}) that there
exists a countable set of bifurcation curves for
$\Omega_1 > \Omega_2$, because $L$ is a positive operator and $\Omega(r)$ is strictly positive. On the
other hand, in the quadrant $\Omega_1 < 0$ and $\Omega_2 > 0$,
there exists another set of bifurcation curves, because $\Omega$ is sign-indefinite and
$L$ is unbounded.

\begin{figure}
\includegraphics[width=0.5\textwidth]{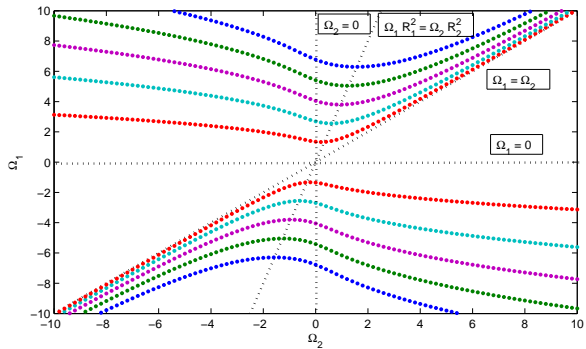}
\caption{Curves of zero eigenvalues (the stability domain is located between red curves)
for $R_1 = 1$, $R_2 = 2$, $b_0 = 0.4$, and $k = 1$. The curves
approach the line $\Omega_1 = \Omega_2$ as $b_0 \to 0$.}
\label{Figure3}
\end{figure}

To study further the instability boundaries, we notice that $\Omega(r)$ depends on both $\Omega_1$ and $\Omega_2$.
Therefore, we shall
rewrite (\ref{eigenvalueCVzero}) as the quadratic eigenvalue problem with the new
eigenvalue parameter $c$ in (\ref{constants-a-b}),
\begin{equation}
\label{stability-curve}
b_0^2 (k^2 + L) U_r = \frac{4 \Omega_2}{r^2} c U_r + \frac{4}{r^2} \left( \frac{1}{r^2} - \frac{1}{R_2^2}\right) c^2 U_r.
\end{equation}

Figure \ref{Figure3} shows numerical approximations of the first five curves of zero eigenvalues
in the upper half of the $(\Omega_1,\Omega_2)$-plane
for fixed values of $R_1 = 1$, $R_2 = 2$, $b_0 = 0.4$, and $k = 1$ and their mirror reflections in the lower half plane. The dotted curves show
the diagonal line $\Omega_1 = \Omega_2$, the Rayleigh line $\Omega_1 R_1^2 = \Omega_2 R_2^2$, as well as the axes
$\Omega_1 = 0$ and $\Omega_2 = 0$.
It is clear that each curve approaches the diagonal line $\Omega_1 = \Omega_2$ for large values of
$\Omega_{1,2}$. When $b_0$ becomes small, they approach closely to the line $\Omega_1 = \Omega_2$.

The above conclusions also
follow from rigorous analysis of the quadratic eigenvalue problem (\ref{stability-curve}).
In the limit $\Omega_2 \to \infty$, we can set $\lambda = \Omega_2 c$ as a new eigenvalue
and treat the last term in (\ref{stability-curve}) as a small bounded perturbation to the
unbounded operator. In the limit $b_0 \to 0$, we set $c = b_0^2 \lambda$ and again treat
the last term in (\ref{stability-curve}) as a small perturbation. In both cases,
eigenvalues $\lambda$ approach to the first eigenvalues of the positive unbounded operator $r^2 (k^2 + L)$.
We note, however, that this approximation is not uniform for all bifurcation curves and only
apply to the finitely many bifurcation curves.

To summarize, we gave mathematically rigorous proofs about distributions and bifurcations of eigenvalues
of linearized operators associated with an ideal hydromagnetic CT-flow
that lay a firm basis for identification of unstable modes in MRI experiments with real dissipative liquids.

\begin{acknowledgments}
O.K. thanks Frank Stefani for fruitful discussions. D.P. is supported by
the AvH Foundation. G.S. is supported by DFG through the Excellence cluster SimTech.
\end{acknowledgments}


\begin{thebibliography}{30}

\bibitem{CI94}
P. Chossat, G. Iooss, {\em The Couette-Taylor Problem},
Springer, New-York (1994); R. Tagg,
{Nonlin. Sci. Today}, \textbf{4}(3), 1 (1994).

\bibitem{r17}
J.W.S. Rayleigh,
Proc. R. Soc. Lond. A. \textbf{93}, 148 (1917).

\bibitem{t23}
G.I. Taylor, Phil. Trans. R. Soc. Lond. A \textbf{223}, 289 (1923).

\bibitem{lathrop04}
D.R. Sisan et al., Phys. Rev. Lett. \textbf{93}, 114502 (2004);
F. Stefani, et al., Phys. Rev. Lett. \textbf{97}, 184502 (2006);
F. Stefani, A. Gailitis, G. Gerbeth, ZAMM \textbf{88}(12), 930 (2008);
M.D. Nornberg et al.,
%M.D. Nornberg, H. Ji, E. Schartman, A. Roach, J. Goodman,
Phys. Rev. Lett. \textbf{104}(7), 074501, (2010);
H. Ji, Proc. Intern. Astron. Union, \textbf{6}, 18 (2010);
M.S. Paoletti, D.P. Lathrop,
Phys. Rev. Lett. \textbf{106}, 024501 (2011);
S. Balbus, Nature \textbf{470}, 475 (2011).

\bibitem{bh91}
S.A. Balbus, J.F. Hawley,
Astrophys. J. \textbf{376}, 214 (1991);
S.A. Balbus, J.F. Hawley,
Rev. Mod. Phys. \textbf{70}, 1 (1998).

\bibitem{v59}
E.P. Velikhov,
Sov. Phys. JETP-USSR \textbf{9}(5), 995 (1959).

\bibitem{c60}
S. Chandrasekhar,
PNAS \textbf{46}, 253 (1960).

\bibitem{DiPrima84}
Th. Gebhardt, S. Grossmann, Z. f. Phys. B. \textbf{90}, 475 (1993);
G. R\"udiger, Y. Zhang, {A \& A}, \textbf{378}, 302 (2001);
H. Ji, J. Goodman, A. Kageyama,
MNRAS \textbf{325}, L1 (2001);
A.P. Willis, C.F. Barenghi,
{A \& A}, {\bf 388}, 688 (2002);
I. Herron, Anal. Appl., \textbf{2}, 145 (2004);
B. Dubrulle et al. Phys. Fluids \textbf{17}, 095103 (2005).

\bibitem{ah73}
D.J. Acheson, R. Hide, Rep. Progr. Phys. \textbf{36}, 159 (1973);
S.A. Balbus, Ann. Rev. Astron. Astroph., \textbf{41}, 555 (2003);
E.P. Velikhov, JETP Letters, \textbf{82}(11), 690 (2005);
D.A. Shalybkov, Physics-Uspekhi, {\bf 52}(9), 915 (2009).


\bibitem{ks11}
O.N. Kirillov, F. Stefani, Astrophys. J. {\bf 712}, 52 (2010);
Phys. Rev. E. (2011) (in press) arXiv:1104.0677

\bibitem{s33}
J.L. Synge, Trans. R. Soc. Can. \textbf{27}, 1 (1933);
P.G. Drazin, W.H. Reid,
\emph{Hydrodynamic stability},
Cambridge Univ. Press, Cambridge, UK (1981).



\bibitem{eva08}
E. M. Graefe et al., J. Phys. A: Math. Theor. \textbf{41}, 255206, (2008);
%E. M. Graefe, U. Gunther, H. J. Korsch, A. E. Niederle, J. Phys. A: Math. Theor. \textbf{41}, 255206, (2008);
%E. M. Graefe, H. J. Korsch, and A. E. Niederle, Phys. Rev. Lett. \textbf{101}, 150408 (2008);
A. A. Sukhorukov, Z. Xu, Yu. S. Kivshar, Phys. Rev. A \textbf{82}, 043818 (2010).
K. Li, P. G. Kevrekidis, Phys. Rev. E \textbf{83}, 066608 (2011).

\bibitem{kcg02}
R. Keppens, F. Casse, and J. P. Goedbloed,
Astrophys. J., \textbf{569}, L121 (2002).

%\bibitem{DiPrima84}
%R.C. DiPrima, P. Hall, Proc.  R. Soc. Lond. A \textbf{396}, 75 (1984);
%K. Julien, E. Knobloch, Phil. Trans. R. Soc. A., \textbf{368}, 1607 (2010);
%I. Herron, J. Goodman, ZAMP, \textbf{61}, 663 (2010).

\bibitem{Kollar} R. Kollar, SIAM J. Math. Anal. {\bf 43}(2), 612 (2011).

%\bibitem{kv10}
%R. Krechetnikov, J.E. Marsden,
%Rev. Mod. Phys. \textbf{79}, 519 (2007);
%O.N. Kirillov, F. Verhulst,
%Z. Angew. Math. Mech. \textbf{90}, 462 (2010).

%\bibitem{gj02}
%J. Goodman, H. Ji,
%\textit{J. Fluid Mech.}, \textbf{462}, 365 (2002).

%\bibitem{rss03}
%G. R\"udiger, M. Schultz, D. Shalybkov,
%\textit{Phys. Rev. E.}, {\bf 67}, 046312, (2003).
\end{thebibliography}
\end{document}